\documentclass[12pt]{article}

\usepackage{graphicx}
\graphicspath{{./figures/}}
\usepackage{dcolumn}
\usepackage{bm}
\usepackage{amsfonts}
\usepackage{amsthm}
\usepackage{amsmath}
\usepackage{amssymb}
\usepackage{slashed}
\usepackage{color}
\usepackage{textcomp}
\usepackage{subfig}
\DeclareMathOperator{\tr}{tr}
\usepackage{hyperref}

\def\be{\begin{equation}}
\def\ee{\end{equation}}
\def\ba{\begin{eqnarray}}
\def\ea{\end{eqnarray}}

\def\bdm{\begin{displaymath}}
\def\edm{\end{displaymath}}
\def\la{~\mbox{\raisebox{-.6ex}{$\stackrel{<}{\sim}$}}~}
\def\ga{~\mbox{\raisebox{-.6ex}{$\stackrel{>}{\sim}$}}~}
\def\bq{\begin{quote}}
\def\eq{\end{quote}}

 at 10truept

\newcommand{\de}{\partial}
\newcommand{\rmd}{\mathrm{d}}
\newcommand{\nab}{\nabla}

\renewcommand{\[}{\left[}
\renewcommand{\]}{\right]}
\renewcommand{\(}{\left(}
\renewcommand{\)}{\right)}

\newcommand{\eps}{\epsilon}

\renewcommand{\th}{\theta}
\newcommand{\lam}{\lambda}
\newcommand{\Lam}{\Lambda}

\newcommand{\vphi}{\varphi}

\newcommand{\Om}{\Omega}

\newcommand{\eV}{\textrm{eV}}

\newcommand{\Mpl}{M_{\textrm{Pl}}}

\newcommand{\bea}{\begin{eqnarray}}
\newcommand{\eea}{\end{eqnarray}}

\newcommand{\bi}{\begin{itemize}}
\newcommand{\ei}{\end{itemize}}

\newcommand{\bfk}{{\bf k}}

\newcommand{\beq}{\begin{equation}}
\newcommand{\eeq}{\end{equation}}
\newcommand{\beqa}{\begin{eqnarray}}
\newcommand{\eeqa}{\end{eqnarray}}

\def\la{~\mbox{\raisebox{-.6ex}{$\stackrel{<}{\sim}$}}~}
\def\ga{~\mbox{\raisebox{-.6ex}{$\stackrel{>}{\sim}$}}~}

\def\ltap{\ \raise.3ex\hbox{$<$\kern-.75em\lower1ex\hbox{$\sim$}}\ }
\def\gtap{\ \raise.3ex\hbox{$>$\kern-.75em\lower1ex\hbox{$\sim$}}\ }
\def\gl{\ \raise.5ex\hbox{$>$}\kern-.8em\lower.5ex\hbox{$<$}\ }
\def\roughly#1{\raise.3ex\hbox{$#1$\kern-.75em\lower1ex\hbox{$\sim$}}}

\begin{document}

\thispagestyle{empty}
\begin{flushright}
May 2016 \\
CERN-TH-2016-104
\end{flushright}
\vspace*{1.25cm}
\begin{center}
{\Large \bf Quantum Field Theory of Interacting Dark}\\
\vskip.3cm
{\Large \bf Matter/Dark Energy: Dark Monodromies}\\

\vspace*{1.25cm} {\large Guido D'Amico$^{a, }$\footnote{\tt
damico.guido@gmail.com}, Teresa Hamill$^{b, }$\footnote{\tt
teresahamill@gmail.com} and Nemanja Kaloper$^{b, }$\footnote{\tt
kaloper@physics.ucdavis.edu}}\\
\vspace{.3cm} {\em $^a$Theoretical Physics Department, 
CERN, Geneva, Switzerland}\\
\vspace{.3cm} 
{\em $^b$Department of Physics, University of
California, Davis, CA 95616, USA}\\

\vspace{1.5cm} ABSTRACT
\end{center}
We discuss how to formulate a {\it quantum field theory} of dark energy interacting with dark matter. We show that the proposals based on the assumption that dark matter is made up of heavy particles with masses which are very sensitive to the value of dark energy are strongly constrained. Quintessence-generated long range forces and radiative stability of the quintessence potential require that such dark matter and dark energy are completely decoupled. However, if dark energy and a fraction of dark matter are very light axions, they can have significant mixings which are
radiatively stable and perfectly consistent with quantum field theory. Such models can naturally occur in multi-axion realizations of monodromies. The mixings yield interesting signatures which are observable and are within current cosmological limits but could be constrained further by future observations.

\vfill \setcounter{page}{0} \setcounter{footnote}{0}
\newpage

\section{Introduction}

Close to $95\%$ of the Universe is invisible. About a quarter of it is compressible, behaving as dark matter (DM). The remaining three quarters have negative pressure, are called dark energy (DE), and approximate the cosmological constant. To leading order these two components seem to be mutually non-interacting in the absence of gravity. The origin of their scales and their relative normalization remains mysterious, particularly in light of the cosmological constant problem \cite{zeldovich,wilczek,wein,kalpad}. While we expect to observe and study systematically the dark matter, DE is extremely difficult to explore beyond simply establishing its existence.

Whatever we think about the cosmological constant problem, nature solves it in some way. So one might wonder if the residual components of the dark sector might be some relics of the mechanism which stabilizes the vacuum energy. This might suggest that the dark degrees of freedom are not totally decoupled from each other. They might have interactions which transfer energy from one dark sector to another. This question is particularly appealing since such energy rebalances might be probed with direct observations of large scale structure and CMB in the forthcoming experiments. 

It is easy to imagine DM/DE mixing \cite{peebles}, especially since one might resort to a fluid approximation to describe their influence on astronomical scales \cite{wetterich}-\cite{ed}. The question is whether such setups make sense in a realistic quantum field theory. One might be tempted to ignore this issue, arguing that we don't really know much about dark sectors. For example, some question whether QFT even works as usual in the dark sector, since after all it suffers from the cosmological constant problem. However, 
QFT has proven time and again to be the most reliable tool for the description of consistent interacting models in nature. So our starting point here will be to seek for a QFT description of interacting DM/DE. 

We will outline the key obstructions to formulating an interacting DM/DE theory in QFT both conceptually and phenomenologically. The main conceptual obstacle stems from {\it decoupling}: one expects that DM should be heavy fields, and DE - if dynamical - light. Because of decoupling, in a normal perturbative QFT very heavy and very light fields do not interact very efficiently. If one introduces strong couplings by hand, these lead to the renormalization of the dark energy scales by the exchange of dark matter quanta. That would render the DE potential very steep, and so DE could not be a field in slow roll any more. The only way for such a potential to behave as DE is if the field is in the minimum with a nonzero cosmological constant. Further, if for any reason the dark energy quanta remain light, their exchange between dark matter particles yields extra long range forces between them, which can easily violate observational bounds on DM interactions.

The combined implications of these problems for interacting DM/DE models are quite severe for a broad class of proposals\footnote{Some bounds have already been considered in the literature, see in particular \cite{khoury,bean}.}. Combining the bounds from the weakness of long range forces and flatness of the  
light DE field potential after DM-induced radiative corrections are included prohibits significant DM/DE mixing when DM are heavy particles with masses $m_{DM} \gg m_{DE}$. 

However, we will show that significant DM/DE mixings are quite likely if a fraction of DM is composed of ultralight boson condensates, with mass not too much greater than the current Hubble scale $H_0$. Thus it is possible to have the mass of this part of the DM close enough to the mass of DE to evade decoupling. 
Using scaled down axion monodromies, which can arise, for example, in the `axiverse' framework \cite{axiverse}, we will see that a system of several
axions can easily yield such a model with dark energy and a fraction of dark matter, with significant DM/DE interactions. We will focus on a set of three axions, of which two are very light, one being DE at the present and another, with a mass $m \ga H_0$, being a component of DM. Since the amount of ultralight dark matter is bounded \cite{barbieri,pedrofco}, in this case most of  DM is provided by a heavier axion\footnote{Or something entirely different, which, as we noted, needs to be decoupled from DE.}, with a mass $m \gg H_0$.
As a result, systems of coupled axions are completely realistic frameworks for interacting DM/DE, and so probing for DM/DE interactions becomes quite sensible. In addition, such tests may be a new portal for exploring the `axiverse.'

\section{Problems with DM/DE Interactions}

So imagine a model of interacting DM/DE within QFT. The observations suggest that at cosmologically large distances any such model should be described by a weakly coupled effective field theory, involving an interaction Lagrangian ${\cal L}_I(\phi, \psi)$. Otherwise it would not even be possible to identify dark matter and dark energy as separate constituents of the universe. Let the field $\phi$ be the DE field (a `quintessence' \cite{wetterich,weiss, steinhardt, kamionaxions}, which is `fundamental' up to some cutoff scale $\mu \gg H_0$) and $\psi$ a DM field. Depending on the mass of the DM field, dark matter is either DM particles (if the mass is large enough so that the particles are nonrelativistic from matter/radiation equality onwards) or an almost uniform, time dependent, condensate of a DM field zero mode (especially if the mass is small; but note that even in this case, $m_{DM} \gg m_{DE}$). Clearly, DM could also be a combination of many separate components.

Suppose first that DM is made up of sufficiently heavy particles. For simplicity\footnote{This is a fairly common assumption \cite{khoury,bean}.}, we could take them to be fermions, with a mass term $m_\psi \bar \psi \psi$. To model DM/DE interactions, let the DM mass depend on the DE field expectation value, $m_\psi = m_\psi(\phi)$. Now, the DE field must not change too quickly, in order to act as dark energy. Further, since we are postulating that the underlying theory of DE is an effective field theory, to ensure its validity we require that $\phi/\Mpl$ is at most of order unity. So we can expand
\be
m_\psi(\phi) = m_\psi^0(1 + c \, \phi/\Mpl + \ldots) =  m_\psi^0+ g \,\phi + \ldots \,. 
\label{mass}
\ee
Here $g = c \, m_\psi^0/\Mpl$, and $c$ is a dimensionless number. So unless $c$ is exactly zero, the DM/DE interactions would involve a Yukawa term
\be
{\cal L} =  g \phi \bar \psi \psi = c \frac{m^0_\psi}{\Mpl} \phi \bar \psi \psi \, . 
\label{yukawa}
\ee
Clearly, prohibiting such a term would require imposing a symmetry in the DE sector. 
\begin{figure*}[thb]
\centering
\includegraphics[scale=0.5]{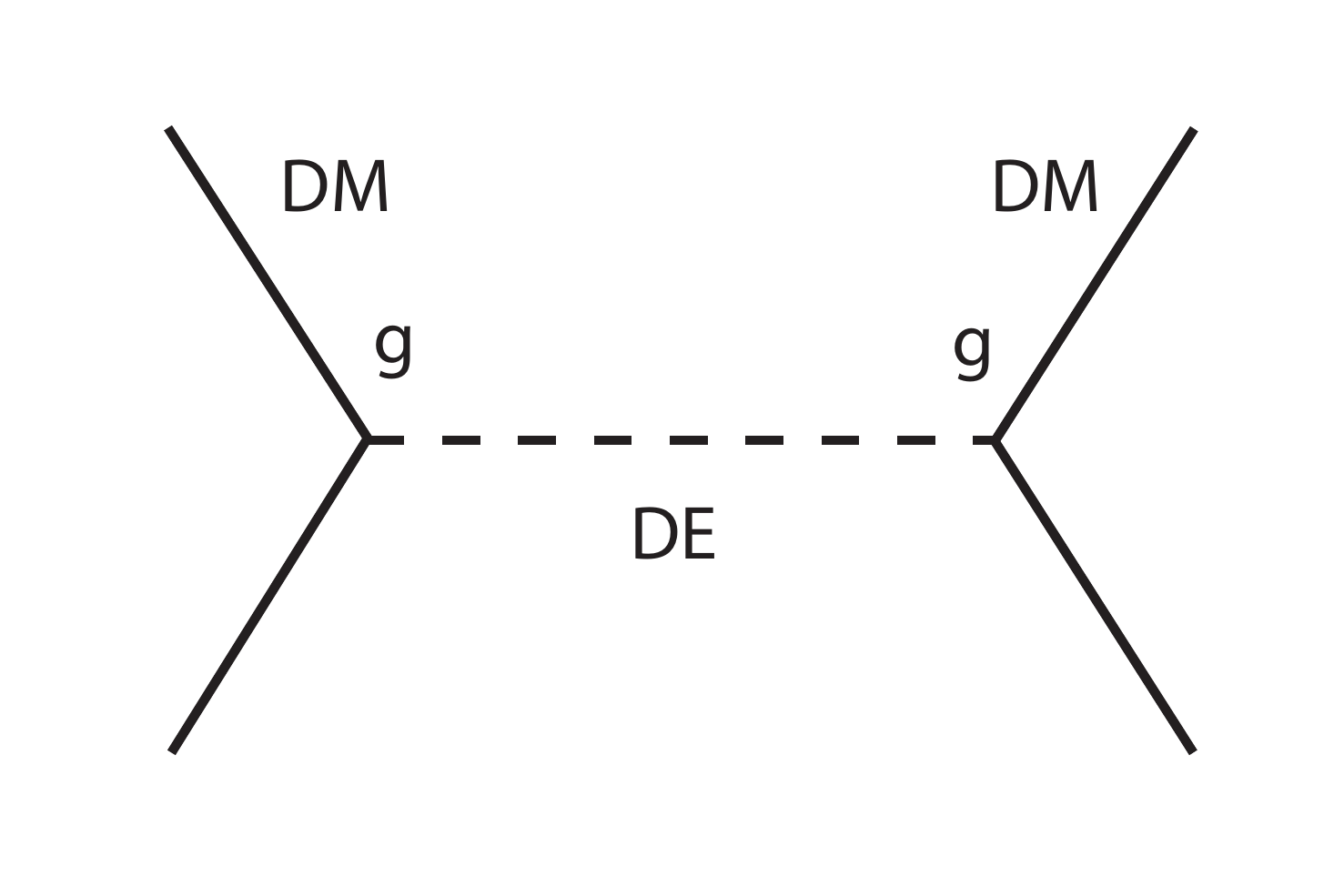}
\caption{Dark matter scattering mediated by dark energy exchange.}
\label{force}
\end{figure*}

In the absence of a symmetry, however, this terms yields new interaction channels. This coupling gives rise to DM/DM interactions depicted by the Feynman diagram of Fig. (\ref{force}).
The exchange of the virtual DE quanta generates an attractive long range force between a pair of DM particles, with an effective potential
\be
V_{DM-DM} \simeq - \frac{g^2}{r} e^{-m_{DE}r} \simeq - c^2 G_N \frac{m_{\psi \, 1}^0 \, m_{\psi \, 2}^0}{r} e^{-m_{DE}r} \, , 
\label{extrapotential}
\ee
where we have used the fact that $1/\Mpl^2 = G_N$. This expression, of course, is just the 3D Fourier transform of the Euclidean DE propagator,
$({\vec p}^{~2} + m_{DE}^2)^{-1}$. Note that while the Yukawa suppression cuts the DE-mediated force off at distances $r \gg 1/m_{DE}$, at shorter distances the resultant force follows the inverse square law.
\begin{figure*}[thb]
\centering
\includegraphics[scale=0.5]{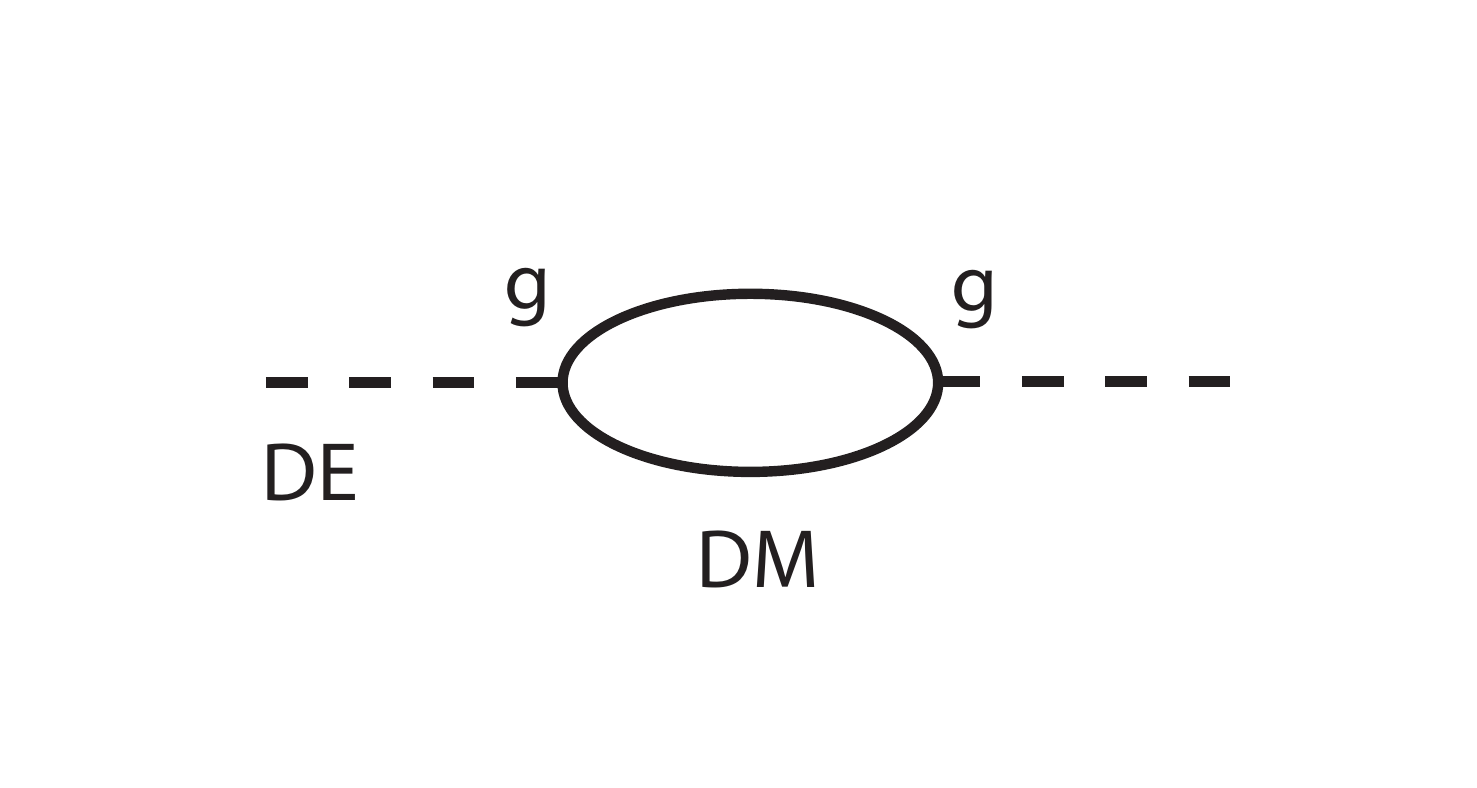}
\caption{Dark energy self-energy from dark matter exchange.}
\label{delmass}
\end{figure*}

Another important consequence of this term arises from the DE self-energy diagram of Fig. (\ref{delmass}), which yields a `dressing' of the DE potential and specifically the DE Lagrangian mass term renormalization due to the emission and reabsorption of the virtual DM quanta. The Lagrangian mass term of the DE field, controlling the steepness of the DE potential, is corrected additively by the contributions from the virtual DM quanta.
The corrections can be calculated using the standard Feynman rules for the diagram (\ref{delmass}), which yield the truncated scalar two-point function
\be
-i \Sigma = g^2 \int \frac{d^4 p}{(2\pi)^4} Tr\Bigl\{\frac{i}{\slashed{p}+\slashed{k} - m_\psi^0}  \frac{i}{\slashed{p} - m_\psi^0}\Bigr\} \, ,
\ee
where $\slashed{k} = \gamma^\mu k_\mu$, and $k$ and $p$ are the external and internal momenta, respectively, and $m_\psi^0$ is the dark matter mass. Rationalizing the integrand by using the Feynman trick to combine the factors in the denominator, performing the trace and analytically continuing to $d=4-\epsilon$ dimensions leads to 
\be
-i \Sigma = 4 g^2 \mu^{4-d} \int \frac{d^d p}{(2\pi)^d} \int_0^1 dx  \, \frac{p^2 - x(1-x)k^2 + m_\psi^0{}^2}{[p^2 + x(1-x) k^2 - m_\psi^0{}^2]^2}\, ,
\ee
where $\mu$ is the regularization scale which preserves the dimensionality of the two point function after continuing to $d=4-\epsilon$ dimensions. 
This integral is divergent, including both divergences which depend on the external momentum and those which don't. The former lead to the
wavefunction renormalization of the scalar, whereas the latter renormalize the mass term in the potential. For our purposes it suffices to focus on the latter, and evaluate the integral at zero external momentum, $k^2=0$. Then after Wick rotation, the dark energy mass term correction is 
\be
\Delta m^2_{DE} 
 = - 4 g^2 \mu^{4-d} \int \frac{d^d p_E}{(2\pi)^d} \, \Bigl\{  \frac{1}{p^2_E + m_\psi^0{}^2} - \frac{2m_\psi^0{}^2}{[p^2_E + m_\psi^0{}^2]^2} \Bigr\} \, .
\label{DEren}
\ee
The integrals are straightforward to evaluate, albeit tedious. We merely state the result here, in the limit when $4 - d = \epsilon \rightarrow 0$:
\be
\Delta m^2_{DE} = \frac{12g^2 m_\psi^0{}^2}{16 \pi^2} \Bigl[ \frac{2}{\epsilon} + \ln\Bigl(\frac{4\pi \mu^2}{{\cal M}^2}\Bigr) - \gamma + \frac13 -
\ln \Bigl(\frac{m_\psi^0{}^2}{{\cal M}^2}\Bigr) \Bigr] \, ,
\ee
where $\gamma$ is the Euler-Mascheroni constant, and the new scale ${\cal M}$ is the renormalization scale, which designates the scale at which the finite part of the renormalized DE mass is determined. The divergent part $\propto 1/\epsilon$ is subtracted by the choice of the bare DE mass counterterm. Using the $\overline{MS}$ subtraction scheme to define the finite part which is cancelled along with the infinity, which means the mass counterterm is
\be
\delta m^2_{DE} = - \frac{12g^2 m_\psi^0{}^2}{16 \pi^2} \Bigl[ \frac{2}{\epsilon} + \ln\Bigl(\frac{4\pi \mu^2}{{\cal M}^2}\Bigr) - \gamma \Bigr] \, ,
\ee
finally leads to the finite mass correction to the DE mass term due to the exchange of virtual DM particles. 

The finite DE mass depends on the physical mass of DM particles,
\be
\Delta m^2_{DE~ {\tt phys}} = \frac{12g^2 m_\psi^0{}^2}{16 \pi^2} \Bigl[ \frac13 -
\ln \Bigl(\frac{m_\psi^0{}^2}{{\cal M}^2}\Bigr) \Bigr] \simeq \frac{3g^2 (m^0_\psi)^2}{2\pi^2}  \ln\(\frac{\cal M}{m^0_\psi}\)  \simeq 
c^2 G_N (m^0_\psi)^4 \ln\(\frac{\cal M}{m^0_\psi}\)\, .
\label{finDEren}
\ee
Multi-loop diagrams involving virtual DM quanta give rise to the additional corrections to the DE Lagrangian mass term, which scale with the same powers of DM mass and the Planck scale. The logarithmic dependence of the correction on DM masses means that it is {\it physical} and cannot be summarily removed at all scales by a single finite renormalization. In other words, although at a fixed scale ${\cal M}$ the corrections to 
$m_{DE}$ could be subtracted away at every order of the loop expansion, the resulting procedure is badly radiatively unstable if the scale of the corrections is much larger than the scale of the observationally required value of the DE Lagrangian mass term. A small change of the renormalization scale ${\cal M}$ would wreak havoc on the cancellation scheme, requiring a completely different prescription for the finite counterterms to maintain the cancellation. This is beside any concern one might have about the UV sensitivity of the theory, which we have completely set aside here.
The point is that the sensitivity of the DE Lagrangian mass term to DM scales is a real thing, as long as the DM particles are much heavier than DE, we insist that the standard rules of QFT apply, and there are no additional symmetries between the scales $m_{DE}$ and $m_{DM}$ to cancel these corrections\footnote{Any additional protection mechanisms that might operate above $m_{DM}$, like SUSY, are irrelevant here.}.

What are the implications of these considerations? First of all, the DE field should be very light to simulate dynamical dark energy - a slowly varying scalar field. Typically this is realized by taking the mass $m_{DE}$ to be smaller than the current value of the Hubble parameter $\lesssim H_0$ for the field to yield at least an e-fold of cosmic acceleration now. Hence, at all subhorizon scales, $r< 1/H_0$, the DE field is effectively massless, giving rise to an additional long range force among dark matter particles, $V_{DM-DM} \simeq - c^2 G_N m^0_{\psi \, 1} m^0_{\psi \, 2}/r$. In the linearized limit, this behaves as a scalar correction to gravity, and its strength relative to the Newtonian potential is
\be
\frac{V_{DM-DM}}{V_N} \simeq c^2 \, .
\label{forces}
\ee
This quantity is constrained by the bounds on the weak equivalence principle violation between dark and ordinary matter \cite{frieman}-\cite{anne} to be less than about $0.1$. Therefore,
\be
|c| \lesssim 1/3 \, .
\label{cbound}
\ee
Note that if we rewrote (\ref{forces}) in terms of the Yukawa coupling $g = c m^0_\psi/\Mpl$, we would get ${V_{DM-DM}}/{V_N} \simeq 
\frac{g^2}{G_N (m^0_{\psi})^2}$, which is essentially the parameter $\beta$ of Eq. (18) of \cite{peebles}, introduced by the comparison of long range dark energy and gravitational fields sourced by dark matter agglomerates, if we ignore the variation of the DE field between the interior and the exterior of the DM agglomerate\footnote{We could have included it here, in which case the results would precisely reproduce those of \cite{peebles}.}. 

Let us now turn to the quantum corrections to DE mass (\ref{finDEren}). As we noted above, DM contributes to the DE Lagrangian mass term regardless of any additional aspects of the UV sensitivity and hierarchy problems. To keep the DE field potential flat despite these corrections, one must require $\Delta m_{DE} \lesssim m_{DE}$. Then (\ref{finDEren}) and $m_{DE} \lesssim H_0$ yield
\be
|c|  \lesssim \frac{\Mpl \, m_{DE}}{(m^0_{\psi})^2} \lesssim \frac{\Mpl \, H_0}{(m^0_{\psi})^2}  \simeq \Bigl(\frac{10^{-3} \, {\rm eV}}{m^0_{\psi}}\Bigr)^2 \, ,
\label{demassbound}
\ee
where in the last equation we substituted numerical values of the Planck scale and the Hubble scale. The point of this equation is to note that as the mass hierarchy between DE and DM increases, the cross coupling, as parameterized by $c$, becomes extremely small quickly. Again, this is perfectly
reasonable since it is but a manifestation of {\it decoupling} of heavy and light modes, which is universally valid in QFT. Moreover, with the actual data reflecting the real world included, equation (\ref{demassbound}) shows that it is generically extremely difficult to couple any DM heavier that a milli-eV to DE without completely destroying the DE sector. We summarize (\ref{cbound}) and (\ref{demassbound}) in Fig. 3.
\begin{figure*}[thb]
\centering
\includegraphics[scale=0.65]{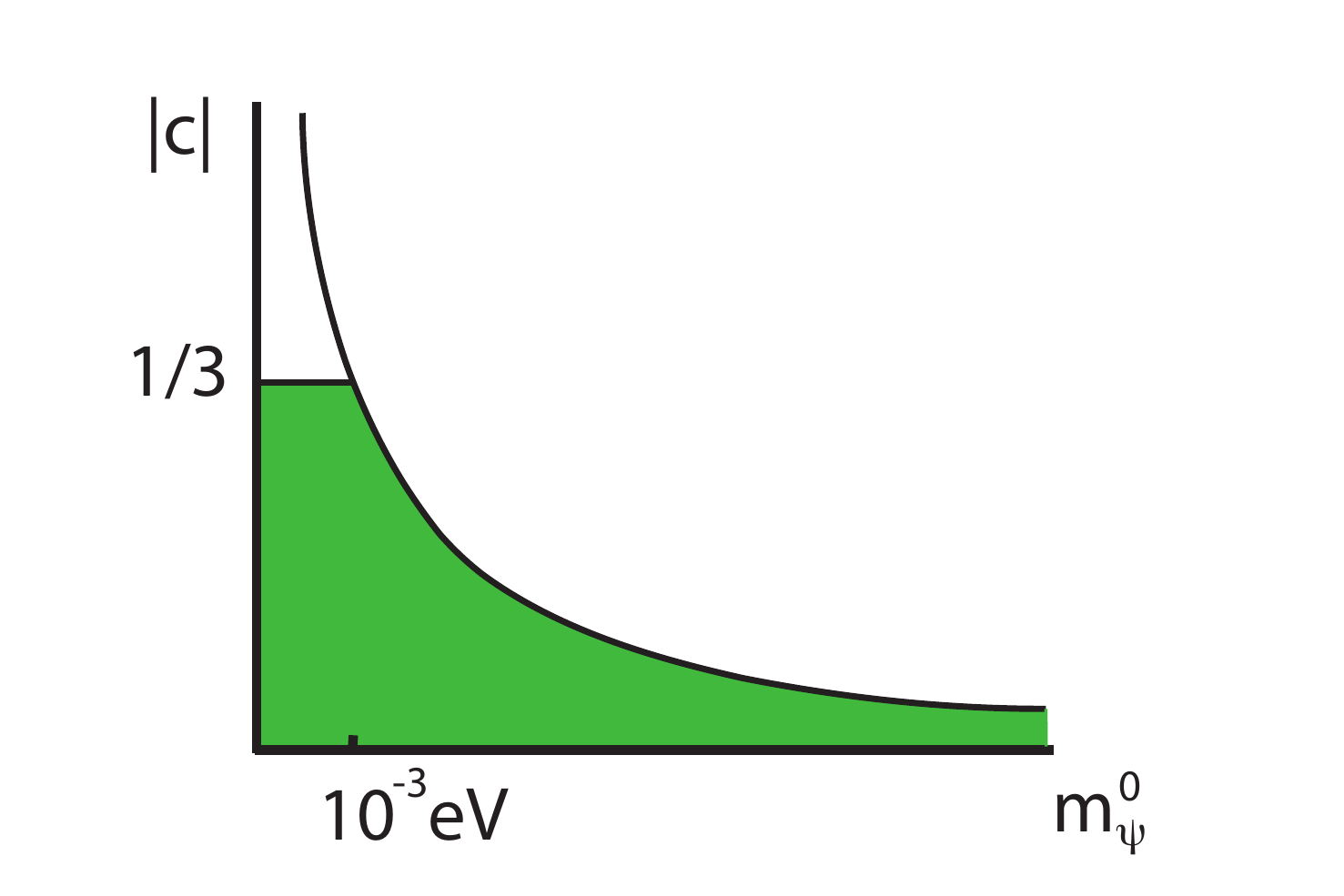}
\caption{DM/DE coupling versus DM mass; shaded area below the hyperbola is allowed. Here DE is an ultralight 
quintessence field with a sub-Hubble mass.}
\label{cbounds}
\end{figure*}

So, for standard quintessence fields with masses $\lesssim H_0$, the  DM/DE interactions are extremely suppressed. This renders simple interacting DM/DE models, with large $m_{DM}$ and tiny $m_{DE}$, such as those described in \cite{khoury}, completely unrealistic. The DM/DE interactions must be irrelevant when $m_{DM}$ is large. 
The problem with this, however, is that the phenomenological bounds on DM masses coming from large scale structure 
\cite{dodelson} require that most of DM is considerably heavier than $10^{-3} {\rm eV}$. Hence, if there are very light `dark' degrees of freedom, which can interact significantly with quintessence, they cannot constitute much of the cosmic contents. As a result, they cannot affect DE evolution very much. Note that while we have used fermions as an example of DM, we would get exactly the same bounds if DM were bosonic particles with trilinear couplings to DE. Note also that we have taken here DE to be a scalar, as is clear from the fermion mass term $\propto \bar \psi \psi$ in (\ref{yukawa}). If $\phi$ had been pseudoscalar, coupling to the fermion bilinear 
 $\propto \bar \psi \gamma^5 \psi$, the bounds from long range forces would essentially disappear since they would be spin-dependent and drop with a higher power of distance \cite{moodywilczek}. The bounds from the radiative corrections would however remain essentially the same as discussed here, still prohibiting the couplings of heavy DM to DE. 

One might try to alleviate this by using a heavier DE field. That could help both with the bounds 
on the corrections to the DE potential and the bounds on the extra long range forces, since the mass corrections could be larger and the force Yukawa-suppressed at distances longer than $1/m_{DE}$. An example is provided by the so-called MaVaN models 
\cite{mavans}. The original proposal assumed that DE is a single scalar with the mass parameter much larger than the Hubble scale, mixing with a visible and sterile neutrino. Although the scalar was too heavy to be in slow roll on its own, the coupling to neutrinos could allow the relic neutrino background to slow down its cosmological evolution. However, this relied on neutrino-DE couplings which were too strong and led to additional attractive long range forces at distances $< 1/m_{DE}$, which, while shorter than the scale of the universe, were still too long range. This forced non-relativistic neutrinos to clump, forming nuggets of size $\sim 1/m_{DE}$, which would not suppress the DE rolling \cite{zaldarriaga}. Indeed, this is in agreement with our calculations of $|c|$
and $V_{DM-DM}$, which show that with $m_{DE} \gg H_0$ one would get a DE-mediated attractive force much stronger than gravity at distances
$\lesssim 1/m_{DE}$. 

Evading this requires either lowering DE mass down to $\lesssim H_0$, or using much more involved DE models, as in the latter of \cite{mavans} (see also \cite{manoj}). These models utilize a hybrid inflation dynamics at extremely low scales, which drives current cosmological acceleration by an effective cosmological constant term (and assumes that all other contributions to the cosmological constant are somehow cancelled). It also includes a coupling of DE to a neutrino species which is always relativistic in order to erase the effects of strong DE-mediated couplings. Yet, in all these cases, most of DM is left completely decoupled from DE. So while the interactions might lead to imprints in the DE evolution, the evolution of DM is not directly affected. 

\section{Axion Monodromies to the Rescue}

The considerations above show that the main obstruction to DM/DE interactions comes from {\it decoupling} in QFT: light and heavy fields do not mix very well. So, this naturally suggests that the strongest interactions occur if DM and DE mass scales are not very different. Now, since DM must be non-relativistic, to accommodate this - i.e. have DM with ultrasmall masses and yet ensure it is not relativistic - one needs DM to be a condensate like DE. In other words, DM is a spatially homogeneous expectation value of an ultralight scalar field very much like quintessence, but with its mass somewhat larger than the current Hubble scale $H_0$. In this case, instead of being stuck in slow roll, the DM {\it vev} is oscillating about its minimum. To leading order, its energy density behaves just like CDM.

To build models with such fields consistently in QFT, one needs to protect the theory from large radiative corrections, both from the ultralight modes in the dark sector and from any other heavier degrees of freedom. A simple way is to provide the low energy EFT with continuous shift symmetries, which ensure that the ultralight dark modes are only derivatively coupled in perturbation theory. Hence their potentials are automatically very flat and radiatively stable. In fact, the only fully self-consistent models of dark energy rely precisely on constructions involving pseudo-Nambu-Goldstone  bosons as DE fields (pNGBs, or `axions') \cite{fhsw}-\cite{kalsor}, which utilize shift symmetry as the protection mechanism of the effective theory. So in practice all one needs is to add an extra axion, over and above the one playing the role of DE, and pick its scales so that it can be DM today \cite{barbieri,gruzinov,jiji}. However, if it is very light, this DM axion is constrained by cosmological data and cannot be all of DM today \cite{barbieri} (see also \cite{pedrofco} for the updated bounds). Even so, it could comprise as much as $\sim 5 \%$ of total amount of DM, which is a far larger fraction of critical energy density than could be in, for example, sterile neutrinos and other light fermions. 

This reasoning points directly to a possible framework for realizing such a scenario\footnote{Note that such a scenario can also occur accidentally, if there are several ultralight axions in nature. The `axiverse' makes the presence of a plethora of very small scales appear more natural.}: the `axiverse' \cite{axiverse}. Multiple axions, with varying scales which control their couplings and masses, arise from string compacifications. The vast diversity of the multiverse and the proliferation of the pNGB fields suggests that such hierarchical axion models are readily realized. In such models the axions are naturally mixed, which means that in low energy EFT they couple to each other \cite{kalsor2,trivedi}. In fact, the simplest mechanisms to generate small axion masses based on axionic monodromies \cite{peloso,evas,flux} automatically ensure that the interactions between the lightest axions are so strong that the axion mixing energy is comparable to the energies in the normal modes. Hence such models can leave quite strong cosmological signatures, affecting directly both DE and DM evolution, by rebalancing the energy contents of individual channels in the course of cosmic evolution.

In what follows we will analyze a simple model, which is essentially a combination of the ideas of \cite{barbieri} and \cite{kalsor2,trivedi,peloso} to demonstrate this. Ultimately, we will be led to a setup with three axions where one is heavy, another is lighter than the current Hubble scale $H_0$, and the third has a mass in between but relatively close to $H_0$. By decoupling, the interactions between the heaviest and the two light axions are small, and so we will ignore them completely. On the other hand, we will devise a simple setup where two light axions mix strongly, so as to ensure that the axion decay constants are not super-Planckian, while one still happens to be lighter than $H_0$. This sector is basically a scaled down version of misaligned axion inflation of \cite{peloso,kkls}. The underlying monodromy which generates ultralight axions in slow roll with sub-Planckian axion decay constants implies that the two lightest axions
automatically have strong interactions. We will analyze the cosmological evolution of the system, both on the cosmological background and with perturbations, to identify the possible signatures and show that the model is consistent with the current bounds.

\subsection{Decoupled Limit}

To set the stage for the perturbative description of the multi-axion dark sector, we start with the case where the interactions are completely turned off \cite{barbieri}. 
So imagine three axions, each with a potential $V_i = \mu^4_i [1 - \cos(\phi_i/f_i)]$. The unity here merely cancels the bare cosmological constant (by hand). The potentials arise from nonperturbative gauge dynamics and are radiatively stable: once flat, they are always flat \cite{fhsw}-\cite{kalsor}, \cite{flux}. The radiative stability of the cosine potentials is ensured by the underlying shift symmetry $\phi_i \rightarrow \phi_i + {\cal C}_i$, broken down to the discrete subgroup
$\phi_i \rightarrow \phi_i + 2\pi N_i f_i$ by the non-perturbative effects, where $N_i$ are integers. The unbroken discrete symmetries ensure that the full nonperturbative potentials are harmonic, periodic functions, where the higher harmonics are suppressed by exponents of $(\Mpl/f_i)^2$. As long as $f_i$'s are not very large the corrections can be ignored \cite{banksdine}. The homogeneous modes' dynamics are 
described by the equations of motion
\be
\ddot{\phi_i} + 3 H \dot{\phi_i} + \partial_{\phi_i} V_i = 0 \, .
\label{decoupledas}
\ee
The fields for which $\partial_{\phi_i} V_i \gg 3 H \dot{\phi_i}$ are DM, with virialized kinetic and potential energy. 
Their homogenous mode oscillates rapidly about the minimum of its potential, with frequency $m_i^2 = \mu^4_i/f^2_i$, and, by the virial theorem, with an amplitude $\phi_i(t) \propto 1/a^{3/2}$. The fields for which 
$\partial_{\phi_i} V_i \ll 3 H \dot{\phi_i}$ are DE, being in slow roll. 

The standard approach to properly normalize the amplitude is to take each light axion to have an initial value $\simeq f_i$ set during
early inflation by the inflationary thermal drift affecting all fields with masses smaller than $H_{inflation}$. Then, for decoupled axions, $f_i$ is the typical distance from the nearest axion minimum, taken as the origin in the field space. This axion stays in slow roll until $H$ drops to $\sim m_i$, and after that starts to oscillate around its nearest minimum, with the amplitude diluting as $\sim 1/a^{3/2}$. If all (or most) of DM today is one such axion, its amplitude at present is given by $m_i^2 \phi_{i, 0}^2 =  3 \Mpl^2 H_0^2 \Om_{m0} \simeq \Om_{m0} h^2 (10^{-3} \eV)^4$. If $f_i$ is fixed, the mass $m_i$ should be picked correctly to match this condition, accounting for the dilution during the evolution of the Hubble scale from $H \sim m_i$ to $H_0$. 

On the other hand, the DE axion is still in slow-roll, frozen at its initial value $\phi_i \simeq O(f_i)$. In this case one needs to have $f_i \simeq \Mpl$, $\mu^4 \simeq 3 \Mpl^2 H_0^2$, in order to match the scale of DE today. Finally, a very light axion which has a mass above $H_0$, but not by much, has been in slow roll until relatively recently. It started to roll around the minimum in a very late universe. This implies that at scales $H>m_{intermediate}$, there was more dark energy than now, and is one of the reasons behind the bounds of \cite{barbieri}. Such an axion didn't dilute too much since it started rolling, but it cannot be more than about $5 \%$ of DM now \cite{barbieri,pedrofco}. \emph{A priori}, its presence seems to require a fair bit of tuning. Yet, as we will see shortly, such tuning can be accommodated rather straightforwardly in monodromies used to generate ultralight mass scales without super-Planckian axion decay constants $f_i$.

\subsection{Mixing and Interactions}

Let us now turn on the interactions between the two lightest axions. We use the effective potential generated by the axions' couplings to dark gauge fields as in a simple monodromy model \cite{peloso,kkls},
\be
V = \mu_1^4 \[ 1 - \cos\(\frac{\phi_1}{f_1} \) \]
+ \mu_2^4 \[ 1 - \cos\(\frac{\phi_2}{f_2} \) \]
+ \mu_3^4 \[1 - \cos\(\frac{\phi_1}{f_1} - n \frac{\phi_2}{f_2} \) \] \, ,
\label{twoaxpot}
\ee
where we again subtract the cosmological constant term by hand, picking $V=0$ at global minima.
We now expand (\ref{twoaxpot}) to quadratic order around a minimum at $\phi_1=\phi_2 = 0$, 
\be
V^{(2)} = \frac{1}{2} \sum_{i,j} \phi_i M_{ij} \phi_j \, ,
\ee
where $M_{ij}$ is the mass matrix
\be
M =
\begin{pmatrix}
\frac{\mu_1^4 + \mu_3^4}{f_1^2} & - n \frac{\mu_3^4}{f_1 f_2} \\
- n \frac{\mu_3^4}{f_1 f_2} & \frac{\mu_2^4 + n^2 \mu_3^4}{f_1^2}
\end{pmatrix}
\equiv
\begin{pmatrix}
m_1^2 & - m_{12}^2 \\
- m_{12}^2 & m_2^2 
\end{pmatrix} \, .
\ee
We can easily transition to the system of normal modes, using a field space rotation by an angle $\tan 2 \th = \frac{2 m_{12}^2}{m_2^2 - m_1^2}$,
which yields the light and heavy axion eigenmodes, respectively,
\be
l = \cos \th \, \phi_1 + \sin \th \, \phi_2 \, , \qquad
h = \cos \th \, \phi_2 - \sin \th \, \phi_1 \, ,
\ee
with eigenvalues $\lam_{\pm} = [ m_1^2 + m_2^2 \pm \sqrt{(m_2^2 - m_1^2)^2 + 4 m_{12}^4} ]/2$. The light field $l$ 
is DE, and the heavy one $h$ is a DM component. This requires $\sqrt{\lam_{-}} \lesssim  10^{-33} \eV$. If $\sqrt{\lam_{+}} \lesssim  10^{-24} \eV$, then its 
abundance is limited to no more than $5\%$ of DM, and depending on the mass, possibly even less. In this case, most of DM has to be something else, e.g. a third axion. On the other hand, if $\sqrt{\lam_{+}} \gtrsim 10^{-24} \eV$, the heavier light axion is not constrained strongly and could be all of DM. For that case, one needs a large separation between the eigenvalues, $\lam_{+}/\lam_{-} \gtrsim 10^{18}$.
In any case, some hierarchy between mass eigenvalues is always needed. 

\subsubsection{Large Hierarchy, Planckian Decay Constants, Decoupling}

To achieve a large hierarchy between the axion masses, we need $\det M \ll (\tr M)^2$, or in terms of the Lagrangian parameters,
\be
f_1^2 f_2^2 (\mu_2^4 \mu_3^4 + \mu_1^4 (\mu_2^4 + n^2 \mu_3^4))
\ll \( f_2^2 (\mu_1^4 + \mu_3^4) + f_1^2 (\mu_2^4 + n^2 \mu_3^4) \)^2 \, .
\ee
In this limit, the eigenvalues become approximately
$\lam_{-} \simeq \frac{\det M}{\tr M}$, $ \lam_{+} \simeq \tr M$, and the mixing angle is tiny, so that 
$l \simeq \phi_1$, $h \simeq \phi_2$.

To simplify the initial analysis, we consider the case when all dimensional parameters are degenerate: $\mu_1 = \mu_2 = \mu_3 = \mu$, $f_1 = f_2 = f_3 = f$, and use the integer $n$ to control the hierarchy. If one light axion is to be quintessence and the other heavy one is to be all of DM, then we need $n \geq 10^9$. Clearly, 
introducing one more axion will relax this. We will consider such cases later on. 

The full potential reads
\be
V = \mu^4 \[ 3 - \cos\(\sqrt{1+n^2} \frac{h}{f}\) - \cos\(\frac{n h + l}{f \sqrt{1+n^2}}\) - \cos\(\frac{n l - h}{f \sqrt{1+n^2}}\) \] \, .
\ee
In terms of the normal modes, $V^{(2)} = \frac{\mu^4}{2 f^2} \[ l^2 + (2+n^2) h^2 \]$, and
the quartic is, to leading order in $n$,
\be
V^{(4)} \simeq - \frac{\mu^4}{24 f^4} \[ n^4 h^4 + \frac{4}{n} h^3 l
+ \frac{12}{n^2} h^2 l^2 - \frac{4}{n} h l^3 + l^4 \] \, .
\ee
Moving away from the degenerate limit, one finds other contributions, and 
in particular, a term $\sim h^3 l$ with a coefficient starting at $O(n)$.

The bottom line, however, is that in this case the interactions are strongly suppressed by the hierarchy parameter $n$.
The cosmological evolution is very close to the one of non-interacting dark matter and dark energy components. This is {\it expected}: 
this is just decoupling in action. 

In this case, the only dimensionless number that controls the interaction between the light and heavy fields is given by the ratio of the masses, and so by the very hierarchy it must be small. So the dark matter is $h$, which will oscillate rapidly, and for sufficiently large $n$ can be all of DM in the universe today. 

The quintessence field $l$, on the other hand, stays in slow-roll.
Since the dimensional coefficients are degenerate, the numerical values will be as in the single-axion case: $\mu \simeq 10^{-3} \eV$, $f \simeq O(\Mpl)$, and the amplitude of $h$ is of order $f/n$. However, this shows that this case does not utilize field space monodromies since 
the axion decay constant $f$ is ${\cal O}(\Mpl)$. All the dynamics unravels in the attractive basin of a single minimum in the theory.

\subsubsection{Less Degeneracy, Sub-Planckian Decay Constants and Monodromies}

More interesting cases can be realized with less degeneracy in the parameter space and by meeting the conditions for monodromy.
To explore this, let us still take the axion decay constants $f_1 = f_2 = f$ for simplicity. Since we imagine that they are determined by 
a symmetry breaking in some UV complete framework such as string theory, this is probably realistic anyway. 
Their magnitude is sub-Planckian, of order 
$\Mpl/S$, with $S \gg 1$ an action of the breaking sector. We will take $f \lesssim 0.01 - 0.1 \Mpl$.
Since the scales $\mu_i$ are generated by non-perturbative physics, they are generically exponentially sensitive to 
$S$, $\mu_i \sim e^{-S_i}$. To illustrate some interesting dynamics, we will focus on $\mu_1 \ll \mu_2 \ll \mu_3$ without loss of generality.
For calculational simplicity, let us further introduce a single parameter $\eps \ll 1$ and 
define $\mu_1 = \eps \mu$, $\mu_2 = \mu$, $\mu_3 = \mu/\eps$. 

The potential now reads (with $V_0$ chosen to  cancel the vacuum energy in the minima)
\be
V = V_0 - \mu^4 \[ \eps^4 \cos\(\frac{\phi_1}{f} \) + 
\cos\(\frac{\phi_2}{f} \) + \frac{1}{\eps^4} \cos\(\frac{\phi_1 - n \phi_2}{f} \) \] \, .
\ee
Expanding in $\eps$ around the vacuum $\phi_1 = \phi_2 = 0$, the eigenvalues are $\lam_{-} \simeq \frac{\mu^4}{f^2 (1+n^2)}$, 
$\lam_{+} \simeq \frac{(1+n^2) \mu^4}{\eps^4 f^2}$. 
The normal modes of the system are $\phi_1 = \frac{n l - h}{\sqrt{1+n^2}}$, $\phi_2 = \frac{n h + l}{\sqrt{1+n^2}}$. In these variables,
\be
V = V_0 - \frac{\mu^4}{\eps^4} \[ \cos\(\sqrt{1+n^2} \frac{h}{f}\) 
+ \eps^4 \cos\(\frac{l + n h}{f \sqrt{1+n^2}}\) + \eps^8 \cos\(\frac{n l - h}{f \sqrt{1+n^2}}\) \] \, .
\label{fullpot}
\ee
Note that when $\epsilon = 1$ this reduces to the previous case. However, having introduced an extra dimensionless parameter $\epsilon$, we can now explore the richer parameter space $(n,1/\epsilon)$ to search for physically interesting regions. 

First off, we see that if we again require $m_h = \mu^2/(\eps^2 f) \gtrsim 10^{-23} \eV$, then the heavy field oscillations can be all of dark matter with mass $m_h$.
To ensure that the other mass is $\le H_0$, we need $\eps \lesssim  10^{-5}$ to get the requisite eigenvalue separation.
This is not all: if the heavy field vacuum is $h=0$, the two fields are essentially decoupled, but the light field $l$ behaves as quintessence only if $f \gtrsim \Mpl$, which we have been trying to avoid in order to guarantee full control over EFT. 

However: if we pick a generic heavy field vacuum $h = 2 k \pi f/\sqrt{1+n^2}$, with $k \gg 1$ integer, keep $f < \Mpl$, and then integrate $h$ out, the effective potential for the light field becomes
\be
V(l) \simeq \frac{\mu^4}{2 f^2 (1+n^2)} \( l + \frac{2 k n \pi f}{\sqrt{1+n^2}} \)^2
+ \eps^4 \mu^4 \[ 1 - \cos\(\frac{2 k \pi}{1+n^2} - \frac{n l}{f \sqrt{1+n^2}} \) \] \, .
\ee
The frequency of the second oscillatory term is very high, and so we do not expand it. It represents merely a small modulation on top of the flat quintessence potential given by the first term, yielding small bumps in dark energy density. Note that the effective light axion decay constant is
$f_{l, \textrm{eff}} \simeq n f$, and so it is ${\cal O}(\Mpl)$ for $n \sim 10-100$ if $f \simeq 0.01 - 0.1 \Mpl$. Further, the light field vacuum is not at zero anymore but at $l_{vac} = 2 k n \pi f/\sqrt{1+n^2} \simeq 2 k \pi f$, which will be super-Planckian for $k > 10$ or so. This is how monodromy sets  up large field excursions from respective vacua.
Essentially, the heavy field pulls the light one's minimum far from the trivial one. Ergo, the slow roll flat plateaus are set up by entirely sub-Planckian local physics. The effective potential remains tiny, of the order $ k^2 \mu^4 / n^2$, even when the field space distance to be traversed is super-Planckian.

For completeness: although DE and DM in this limit remain decoupled, we can easily get an approximate $\Lambda$CDM evolution from entirely sub-Planckian microphysics. To start, we need the equation of state to transition from $w \simeq 0$ at early times towards $-1$ in the future. The oscillations of $h$ around its minimum provide all of the required DM.
Writing $h(t) =  2 k \pi f/\sqrt{1+n^2} + \chi(t)$, the field $\chi(t)$ oscillates around zero with frequency $m_h$ and amplitude decreasing as $a^{-3/2}$.
From the potential above, it is clear that $l$ is in slow roll today as long as $k$ is big enough. 
The resulting energy density and pressure are $\rho \simeq m_h^2 \chi^2 + 2 \mu^4 = \mu^4 ( \frac{1+n^2}{\eps^4} \frac{\chi^2}{f^2} + 2 )$, $
p \simeq - 2 \mu^4$. So, $w_{total} \simeq - 1/ (1+ \frac{1+n^2}{2\eps^4} \frac{\chi^2}{f^2})$ Clearly, this realizes the required limiting behavior: before $\chi$ dilutes enough, so that its energy density dominates, $w_{total} \simeq 0$. As time goes on, $w_{total}$ converges to $-1$ for as long as $l$ remains in slow roll. 
Specifically, under the assumption that the heavier axion is all of DM today, this sets the value of the heavy field displacement 
from the minimum at the present epoch
\be
\chi_0^2 \simeq \frac{\Om_m}{\Om_{\Lam}} \frac{\eps^4 f^2}{1+n^2} \, .
\label{amplitude}
\ee
More generally, this is the upper bound on the value of $\chi$ at the current time.

\subsubsection{Mixing and Monodromies}

The most interesting examples involve a very small hierarchy between the two lightest axions, while still realizing monodromy in the field space. This means that we need the third axion to be most of DM.
To realize this case, we retain the parametrization of scales in the full potential of the previous section, but allow $\epsilon$ to be larger. As we noted previously, the masses of normal modes are $m^2_l \simeq \frac{\mu^4}{f^2(1+n^2)}$ and $m_h^2 \simeq \frac{(1+n^2)\mu^4}{\epsilon^4 f^2}$, so their ratio is $m_l^2/m_h^2 \simeq \epsilon^4/(1+n^2)^2$. Hence taking $\epsilon \lesssim 1$ and $n \sim {\cal O}({\rm few})$ will easily generate a small hierarchy between them. Further, the potential (\ref{fullpot}), which we repeat here,
$$
V = V_0 - \frac{\mu^4}{\eps^4} \[ \cos\(\sqrt{1+n^2} \frac{h}{f}\) 
+ \eps^4 \cos\(\frac{l + n h}{f \sqrt{1+n^2}}\) + \eps^8 \cos\(\frac{n l - h}{f \sqrt{1+n^2}}\) \] \, .
$$

Since most of DM is another, third axion, much heavier than the normal modes $l,h$, we now do not require $\epsilon \ll 1$.  We do require, however, that $m_l \lesssim H_0$ in order for the light mode to be quintessence. Further, as above, we demand that $f< \Mpl$ such that the EFT is under control, including the corrections from quantum gravity. We see that $n \sim {\cal O}({\rm few})$ will help bring the mass $m_l$ down by about an order of magnitude, and combining this with 
$\mu \lesssim 10^{-3} \, {\rm eV}$ will ensure that $m_l \sim H_0$, while $m_h$ is a few orders of magnitude heavier.

Next, we imagine that the heavy field $h$ resides in an attraction basin of a generic vacuum $h_{vac} = 2 k \pi f/\sqrt{1+n^2}$. Shifting the heavy field to $h =  2 k \pi f/\sqrt{1+n^2} + \chi$, we rewrite
\ba
V &=& V_0 - \frac{\mu^4}{\eps^4} \cos\(\sqrt{1+n^2} \frac{\chi}{f}\) 
- \mu^4 \cos\(\frac{l}{f \sqrt{1+n^2}} + \frac{2 k n \pi}{{1+n^2}} + \frac{n\chi}{f \sqrt{1+n^2}}\) \nonumber \\
&-& \mu^4 \eps^4 \cos\(\frac{n l}{f \sqrt{1+n^2}} -\frac{2 k \pi}{{1+n^2}} - \frac{\chi}{f \sqrt{1+n^2}}\) \, .
\label{monopot}
\ea
The first term in (\ref{monopot}) is the potential of the heavy field $h$ - or $\chi$, after the field redefinition. In writing it, we have dropped the phase shift $2 k \pi$. At times when $m_h > H$, this potential forces 
$\chi$ to oscillate around the minimum, contributing to DM. Clearly, the oscillations are bounded, as shown by \cite{barbieri}. This simply means that, at present, the amplitude of $\chi$ cannot be larger than a fraction of (\ref{amplitude}) -- i.e., if the amplitude of $\chi$ is less than a fifth of $\chi_0$ in (\ref{amplitude}), the energy density contribution to DM from the heavier ultralight axion cannot be more than few percent. This would fit the bounds of \cite{barbieri,pedrofco}.

The second term gives the leading order contribution to the light field potential. It features a monodromy. To see this we normalize the argument of the cosine by taking out the light field frequency as a prefactor, so this term is $\propto \cos[\frac{1}{f\sqrt{1+n^2}} (l + \frac{2 k n \pi f}{\sqrt{1+n^2}} + n\chi)]$. The phase shift $ {2 k n \pi}/({{1+n^2}})$ generically cannot be removed by periodicity of the cosine, and for $1 < k < n$, it is $\sim 2 k \pi/n \la 2\pi$. Yet, this automatically induces a displacement of $l$ from its vacuum by ${2 k n\pi f}/{\sqrt{1+n^2}} \simeq 2 k \pi f$.
Even for relatively moderate values of $k$, this makes the effective field excursions of $l$ super-Planckian although $f \lesssim \Mpl$. 
The third term is an additional modulation of the light field potential - as before - producing bumps on the quintessence potential. The reason we are adding it is that, in principle, it helps generate the hierarchy between $l$ and $h$. 

It is important to notice, however, that when $n, 1/\epsilon$ are not extremely large, all the terms in the potential are normalized approximately the same. This means that if we expand (\ref{monopot}) beyond the quadratic order, we will find nonlinear terms, describing interactions between DE ($l$-field) and a component of DM ($h$-field), with energy densities that are comparable in magnitude to the energy density contributions from the normal modes {\it at times before the heavier normal mode dilutes away}. In other words, this guarantees that the DM/DE interactions are significant during at least a brief period in cosmic history. In fact, we will show that the interaction can be strong enough to facilitate a classical transition of the heavier normal mode from one vacuum to a neighboring one, creating a domain wall, whose tensions are fortuitously small enough to meet the observational limits \cite{zeldomain}. In those cases, we find the largest deviations from the $\Lambda$CDM cosmology. In the next section we resort to numerical evolution of the model in this regime to determine the signatures of the interactions. We also consider the effects on cosmological perturbations.

\section{Cosmological Evolution}

To investigate the cosmological signatures of the dynamics, we now turn to numerical analysis. First, we integrate the homogeneous equations to explore the evolution of the background and find out the behavior of the leading order cosmological observables as a function of the redshift $z$. After that, we consider the subleading effects using cosmological perturbation theory. Our focus is on the imprints of DM/DE interactions. We exhibit the similarities -- and crucial and interesting differences -- of the interacting models with monodromy when compared to $\Lambda$CDM and models with ultralight decoupled axions as discussed by \cite{barbieri}. The interacting models may leave nontrivial imprints when compared to these benchmarks, especially when the interactions are strong enough to force a vacuum transitions of the heavier ultralight mode. After that we turn to perturbative analysis of the dynamics and find that interacting models meet the current observational limits.
For the background evolution, we start our integration from high redshift, fixing the parameters to their $\Lambda$CDM values $\Omega_{c0} = 0.26$, $\Omega_{b0} = 0.05$, $\Omega_m = \Omega_c + \Omega_b$, $T_{CMB} = 2.275 {\rm K}$, assuming 3 flavors of massless neutrinos. We normalize the Hubble parameter today to $H_0 = 68 \, {\rm km/s/Mpc}$.
We choose initial conditions for the axion potential such that the Universe is flat, and we set the initial field velocity to zero, as the axions are negligible at the initial epoch. A more complete analysis, including data and a broader allowed range of values of cosmological parameters, would clearly be interesting but is beyond the scope of the present work.

\subsection{Background}

To maximize the mixing between the two lightest axions, we choose parameters that ensure that one of the cosine terms goes over maxima before settling in a minimum\footnote{One can consider the interactions between fields that remain in a single vacuum, but in this case the effect of mixing is diminished and short-lived. This is because once the heavy field settles in its minimum the cosine is well approximated by a quadratic potential, and therefore the mixing term can be set to zero by a field redefinition.}. The vacuum transition is most easily understood using the form of the potential given in Eq. (10):
$$
V = V_0 - \mu_1^4 \cos\(\frac{\phi_1}{f_1} \)
- \mu_2^4 \cos\(\frac{\phi_2}{f_2} \)
- \mu_3^4 \cos\(\frac{\phi_1}{f_1} - n \frac{\phi_2}{f_2} \) \, ,
$$

Let us now take $\mu_2 \gg \mu_1 , \mu_3$ (reordering the scales relative to the previous case) and $f_1 = f_2$. With these parameters, $\phi_2$ begins to oscillate while $\phi_1$ is still in slow roll. If the oscillations of $\phi_2$ are large enough that $n |\Delta\phi_2|/f_2 > 2\pi$, then the third term crosses over a maximum of the cosine. This means that, for given $n$ and $f_2$, the initial value of $\phi_2$ must be a sufficient distance from the minimum of the full potential. This is easily accomplished, even with the constraint that the heavier light axion field is not more than $\sim 5 \%$ of DM today. Further, one should, in principle, also pick the initial conditions for the slowly rolling field so that it rests in a convex section of the cosine, to avoid excessive tachyonic perturbations\footnote{For a short lived stage of late acceleration these bounds are relatively weak.} \cite{kalsor}. These conditions ensure that there will be mixing, with some effect on cosmic expansion. 
The parameters we choose are listed in table~\ref{tab:par_coupled}. For comparison, we also consider a decoupled model where the mixing between the two ultralight axions is completely turned off.  The parameters describing it are listed in table~\ref{tab:par_decoupled}.
\begin{table}[thb]
\centering
\begin{tabular}{ | l | c | c | r | }
\hline  
  $\mu_1^4$ & $1.1 \, \Mpl^2 H_0^2$ & $f_1$ & $0.1 \, \Mpl$ \\
  \hline
  $\mu_2^4$ & $10.75 \, \Mpl^2 H_0^2$ & $f_2$ & $0.1 \, \Mpl$ \\
\hline  
  $\mu_3^4$ & $1.07 \, \Mpl^2 H_0^2$ & $\phi_{1, \textrm{in}}$ & $0.155 \, \Mpl$ \\
\hline
  $n$ & 9 & $\phi_{2, \textrm{in}}$ & $0.7835 \, \Mpl$ \\
\hline
\end{tabular}
\caption{Parameters for the interacting model.}
\label{tab:par_coupled}
\end{table}
\begin{table}[thb]
\centering
\begin{tabular}{ | l | c | c | r | }
\hline  
  $\mu_1^4$ & $ 1.07 \, \Mpl^2 H_0^2$ & $f_1$ & $0.9 \, \Mpl$ \\
  \hline
  $\mu_2^4$ & $6.65 \, \Mpl^2 H_0^2$ & $f_2$ & $0.1 \, \Mpl$ \\
\hline  
  $\mu_3^4$ & $0 $ & $\phi_{1, \textrm{in}}$ & $2.5192 \, \Mpl$ \\
\hline
  $n$ & 0 & $\phi_{2, \textrm{in}}$ & $0.2075 \, \Mpl$ \\
\hline
\end{tabular}
\caption{Parameters for the decoupled model.}
\label{tab:par_decoupled}
\end{table}

\begin{figure*}[h]
\centering
\includegraphics[scale=0.4]{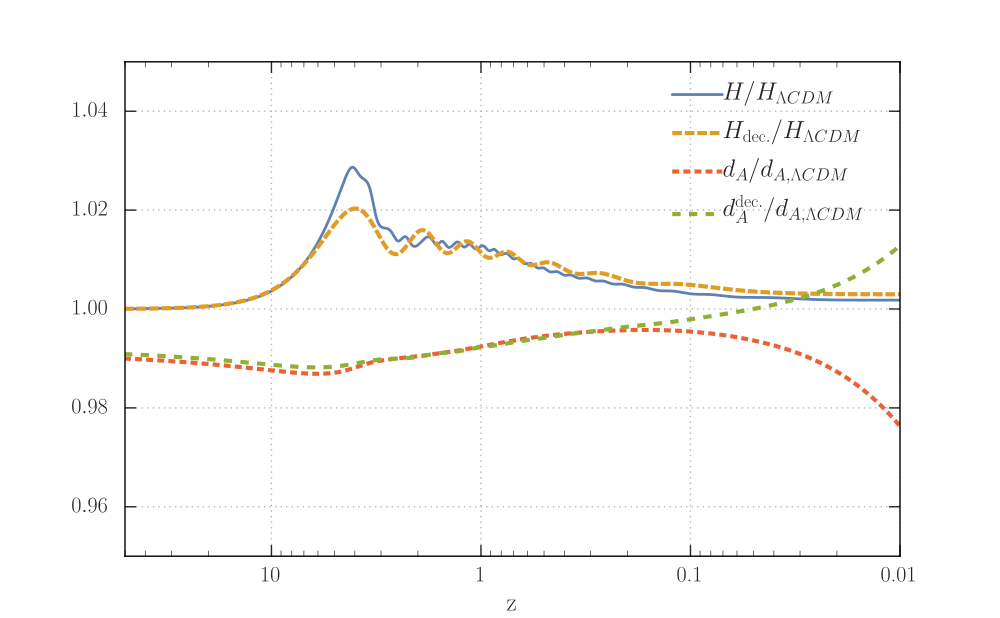}
\caption{Hubble parameter and angular diameter distance compared to the Planck $\Lambda CDM$ value, for the coupled ($H$, $d_A$) and decoupled ($H_{\rm dec}$, $d_A^{\rm dec.}$) models.} 
\label{fig:HdAratios}
\end{figure*}

\begin{figure*}[thb]
\centering
\includegraphics[scale=0.4]{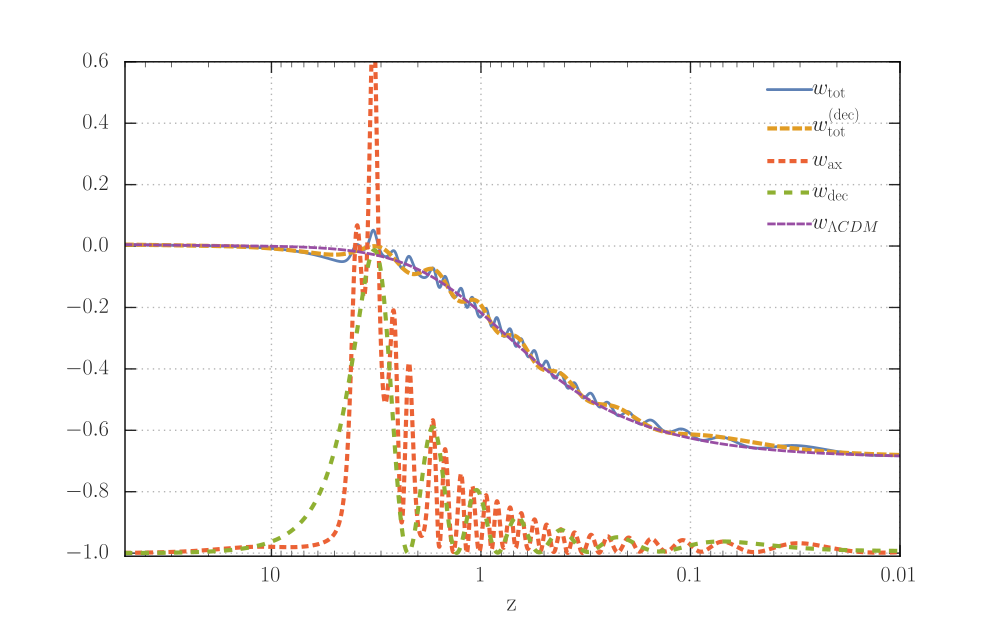}
\caption{The total equation of state and the equation of state of the two light axions, compared to the $\Lambda$CDM case, for both coupled and decoupled models.}
\label{fig:w}
\end{figure*}

Figures (4-7) show the evolution of these models, normalized to $\Lambda$CDM. Both the decoupled and interacting models differ from $\Lambda$CDM by a temporary increase in $H(z)$ at low redshift (and the corresponding variation of $d_A(z)$), Fig.~\ref{fig:HdAratios}. Initially, we normalize $H/H_{\Lambda{\rm CDM}}$ to unity fixing the amount of CDM in the early universe to reflect the data. We add DE at early times in the form of the heavier ultralight axion, which will decay before today. This means that early on there was more DE. The equilibration of the heavier ultralight axion is delayed by its oscillations and also by the interactions.  Nevertheless, asymptotically it removes some fraction of DE and converts it into DM as time goes on, which redshifts away under the influence of DE. In the asymptotic future, this implies that $H/H_{\Lambda{\rm CDM}}$ will dip below unity. In both models, the oscillations of the intermediate-mass axion field produce oscillations in the total equation of state, Fig. (\ref{fig:w}), with corresponding effects in $H$ and $d_A$. Interestingly, these oscillations can mimic a DM/DE interaction even in the decoupled case, when the actual interaction is zero. As the equation of state of the intermediate-mass field oscillates, it goes from being more DE-like to DM-like to DE-like, and so on. 

For a field that has $m\sim O(10)H_0$ these oscillations are slow enough to be observed and do not effectually average out as they do for the main component of dark matter. Essentially, since the mass of the heavier light axion is so close to the Hubble scale, the kinetic and potential energies in the field oscillations do not have time to virialize. Their energy rebalance, as well as the fact that the field just fell out of slow roll, simulate the energy transfer between DE and DM, as if the interactions were present. 
Constraints on axions in this mass regime are discussed in \cite{kamionaxions}, assuming that DE is a single axion.

A scrutiny of the evolution of the equation of state parameter with redshift reveals a striking difference between the coupled and decoupled models,
shown in Fig. (\ref{fig:w}). This happens when the coupled model has interactions strong enough to force a vacuum transition of the heavier light axion. This jump and the subsequent field ringdown is responsible for differences in the shapes of the plots of $H(z)$ and $d_A(z)$. 

While the vacuum transitions are clearly very interesting for inducing the largest allowed deviations from the $\Lambda$CDM evolution, at the same time, they are worrisome since they lead to the formation of domain walls inside the horizon. These domain walls could in principle lead to the large perturbations in the late universe, and are strongly constrained by observations \cite{zeldomain}. To evaluate the constraints, we need to get a reliable estimate of the domain wall tension, which controls the scale of distortions. Since the transition is between nearest neighbor minima in the cosine potential governing the heavier light axion, we can approximate the cosine by the quartic potential, obtained by expanding the cosine about the maximum between the two minima. This yields $V_{eff} \simeq (\frac{\mu_3}{f_{eff}})^4 (h^2 - f_{eff}^2)^2$, where $f_{eff} \simeq f_2/n$ is the effective period of this cosine, as discussed after eq.~\eqref{monopot} above. This means that the field $vev$ in the core of the wall is 
$\sim f_{eff}$, but crucially, the self-coupling is $\lambda_4 \simeq (\mu_3/f_{eff})^4$. Its scaling with four inverse powers of $f_{eff}$ is the reason the domain walls pass the observational bounds. Indeed, the tension of a domain wall separating two nearest neighbor minima is
\be
\sigma \simeq \sqrt{\lambda_4} f_{eff}^3 \simeq \(\frac{\mu_3}{f_{eff}}\)^2 f_{eff}^3 = \mu_3^2 f_{eff} \simeq \frac{\mu_3^2 f_2}{n} \, .
\ee
On average there would be one such domain wall per Hubble volume, and so the fraction of the domain wall energy to the total energy
inside a Hubble patch would be of order
\be
\frac{E_{wall}}{E_{Hubble}} \simeq \frac{\sigma H_0}{\Mpl^2 H_0^2} \simeq \frac{ \mu_3^2 f_2}{n \Mpl^2 H_0} \, .
\ee
Since $\mu_3^2 < \Mpl H_0$ -- as the heavier light axion is just a component of the total energy density of the universe now, and the interactions are a subleading contribution to its energy density --  we find
\be
\frac{E_{wall}}{E_{Hubble}} < \frac{f_2}{n \Mpl} \, .
\label{tensionbound}
\ee
This inequality\footnote{We note that we are somewhat sloppy here, since we are ignoring the numerical prefactor in (\ref{tensionbound}). This prefactor is easily $\lesssim {\cal O}(10^{-2})$ for realistic choices of parameters, and for $n \sim {\cal O}(10)-{\cal O}(100)$ it allows $f_2 \lesssim \Mpl/10$ to satisfy the bound (\ref{tensionbound}). We have also neglected to scale the energy density in the domain walls up by a factor of redshift to the epoch when the walls were produced and compare that to the critical energy density then. Since the relevant factor goes as $\sqrt{z}$, and the redshift when the domain walls are made is $z \la 100$ this is readily compensated by other uncertainties.} can be satisfied with the monodromy configurations by taking $f_2 \ll \Mpl$, which ensures that the domain wall-induced distortions of the cosmological geometry are small enough.
With further expansion of the universe, the energy density in domain walls decreases with respect to dark energy, so its contribution remains subdominant. However, there could be interesting small effects if the bound (\ref{tensionbound}) is close to saturation.

\begin{figure*}[th]
\centering
\includegraphics[scale=0.4]{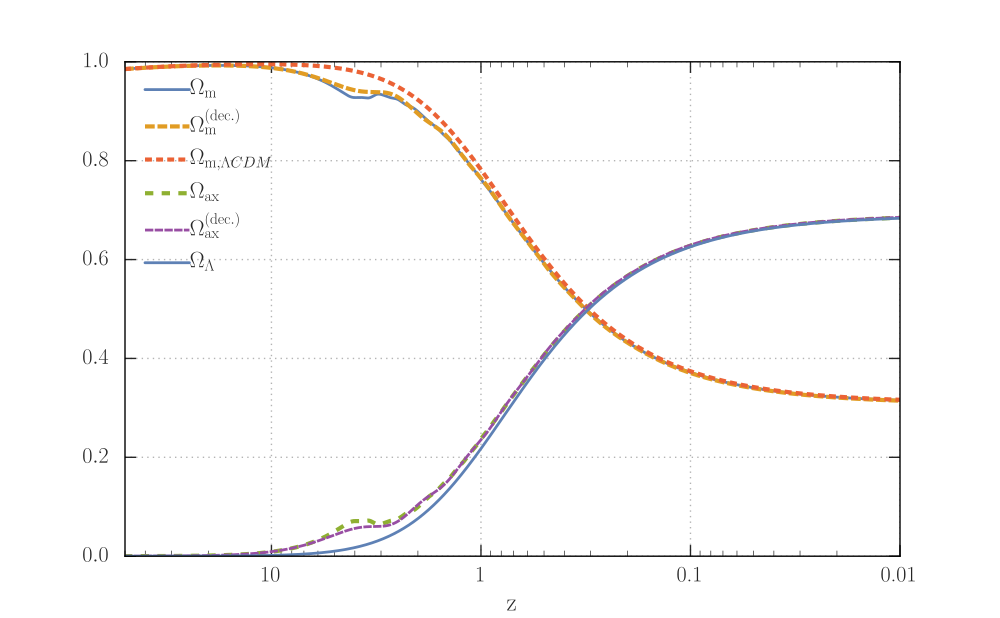}
\caption{Energy densities in the main component of dark matter and in the axion sector, for both coupled and decoupled models.
For comparison, we show the energy density in matter and $\Lambda$ in $\Lambda$CDM.} 
\label{fig:Omega}
\end{figure*}

\begin{figure*}[thb]
\centering
\includegraphics[scale=0.4]{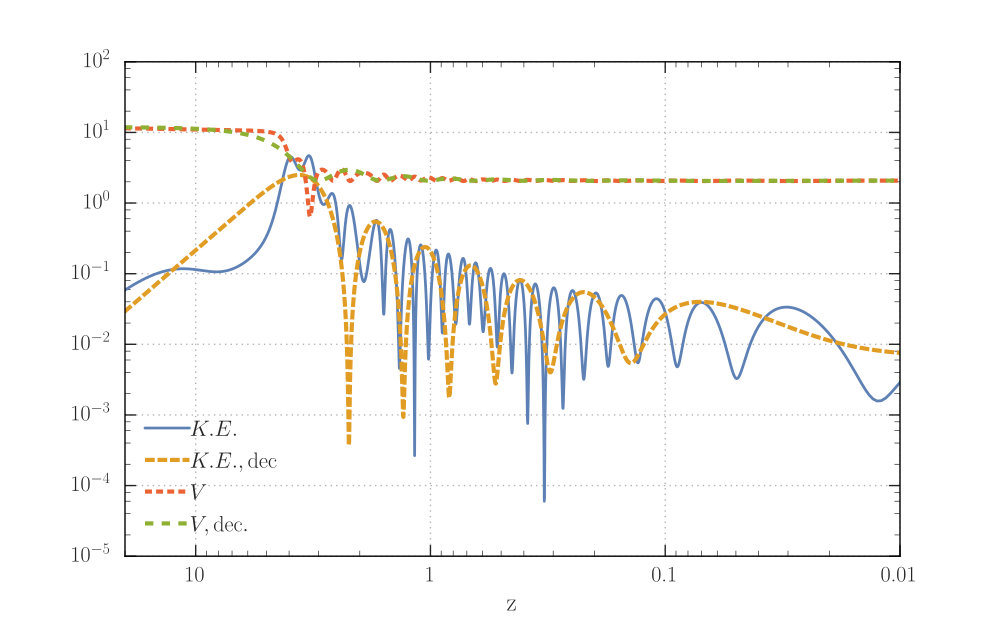}
\caption{Kinetic energy and potential energy in the axion sector.} 
\label{fig:energies}
\end{figure*}

We note that the energy density in the heavier field in the coupled model dissipates away faster than that of simple DM, as displayed in the evolution of the decoupled model, Fig.~(\ref{fig:Omega},\ref{fig:HdAratios}). This is due to the sharp increase in kinetic energy of the intermediate-mass field as it goes over the extrema, as shown in Fig.~(\ref{fig:energies}). During this stage, the kinetic energy dissipates much faster, approaching the $1/a^6$ law. 
This constitutes an important signature of the coupled model: the coupled model may have a comparatively large impact on cosmic expansion at a particular redshift and a comparatively small effect at later times.\footnote{This will modify the bounds coming from \cite{ barbieri,pedrofco}.} 
Finally, we note that the fluctuations about the background satisfy a redshift-dependent dispersion relation. This is because the effective mass of the intermediate-mass axion is background dependent due to the interaction with the DE axion.

\subsection{Perturbations}

Finally, we turn to the linearized cosmological perturbations of the models with two light axions and cold dark matter. Our purpose is not a detailed study of the perturbation effects, i.e., a full analysis that would be needed for the detailed comparison with data. We merely want to demonstrate the consistency of the axion DM/DE models with the current bounds, by comparing their late time evolution to the standard $\Lambda$CDM model. This is the key consistency test, since at early times the matter contents and behavior is approximately the same in both models. We stress that some aspects of this have already been done. Specifically, an extensive analysis of the effects of a single ultra-light axion, which is a limit of the decoupled model with two ultralight axions we have been using here, has been done in~\cite{barbieri, pedrofco}. A similar analysis of the coupled dynamics with nontrivial mixing would therefore seem very warranted but is left for future work.  

\begin{figure*}[thb]
\centering
\subfloat{\includegraphics[width=0.49\textwidth]{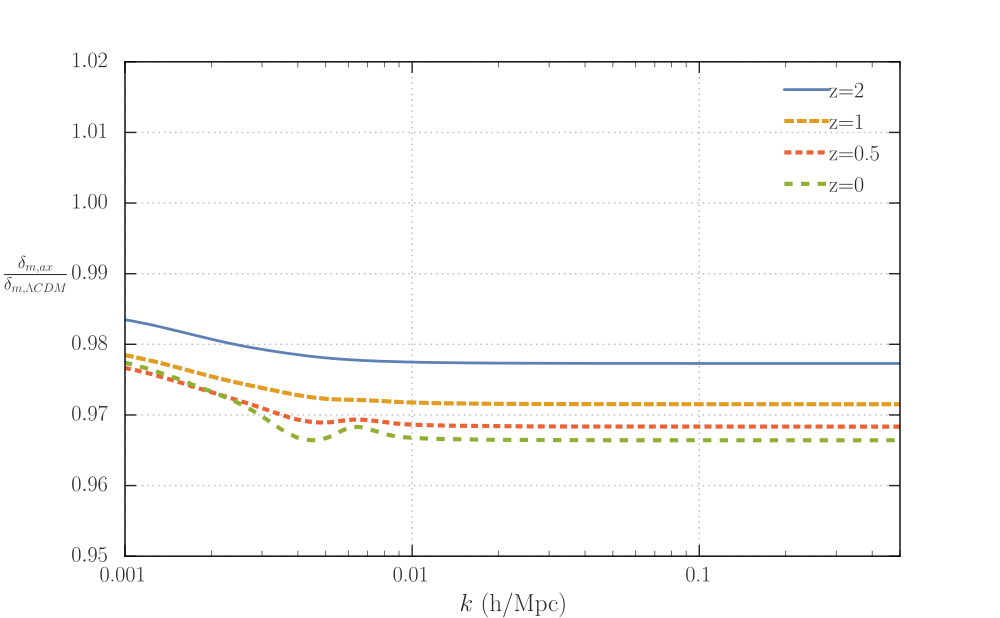}}
\subfloat{\includegraphics[width=0.49\textwidth]{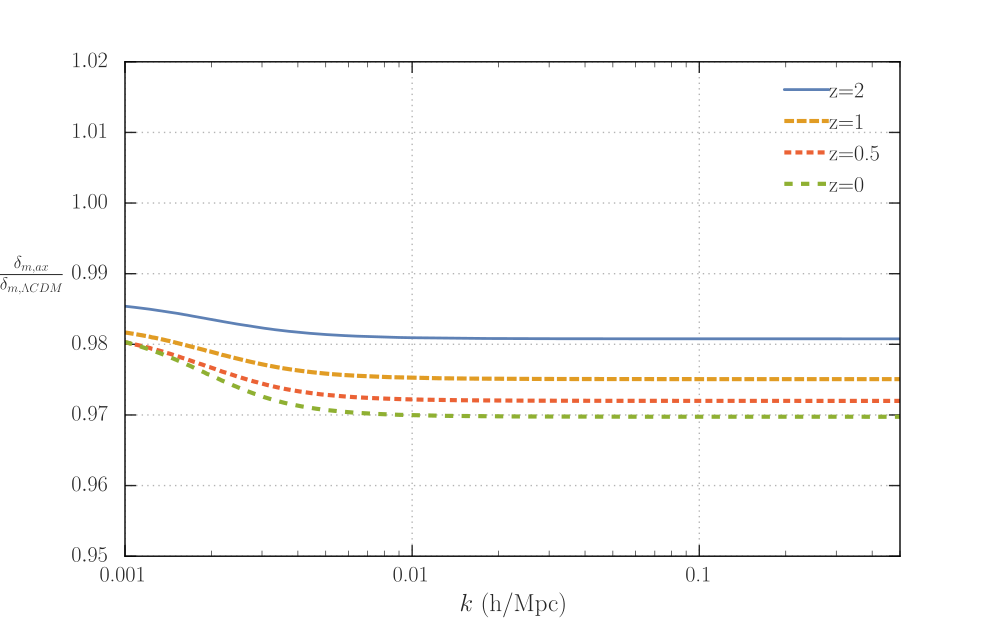}}
\caption{Ratio of axion matter perturbations to the $\Lambda$CDM matter perturbations for redshifts 2, 1, 0.5 and 0.
{\it Left}: interacting model; {\it Right}: decoupled model.}
\label{fig:deltam}
\end{figure*}

\begin{figure*}[ht]
\centering
\subfloat{\includegraphics[width=0.49\textwidth]{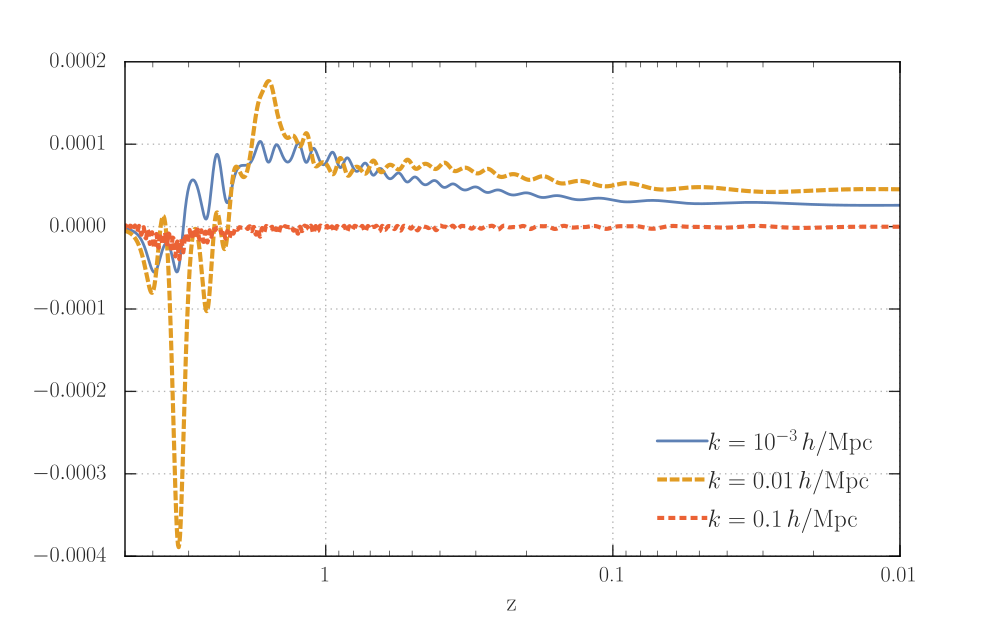}}
\subfloat{\includegraphics[width=0.49\textwidth]{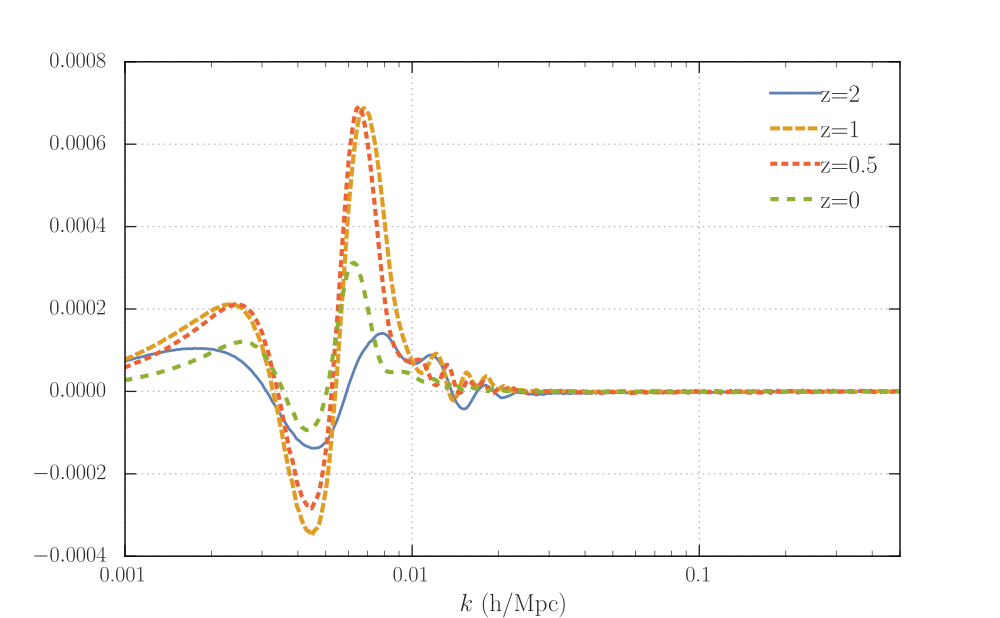}} \\
\subfloat{\includegraphics[width=0.49\textwidth]{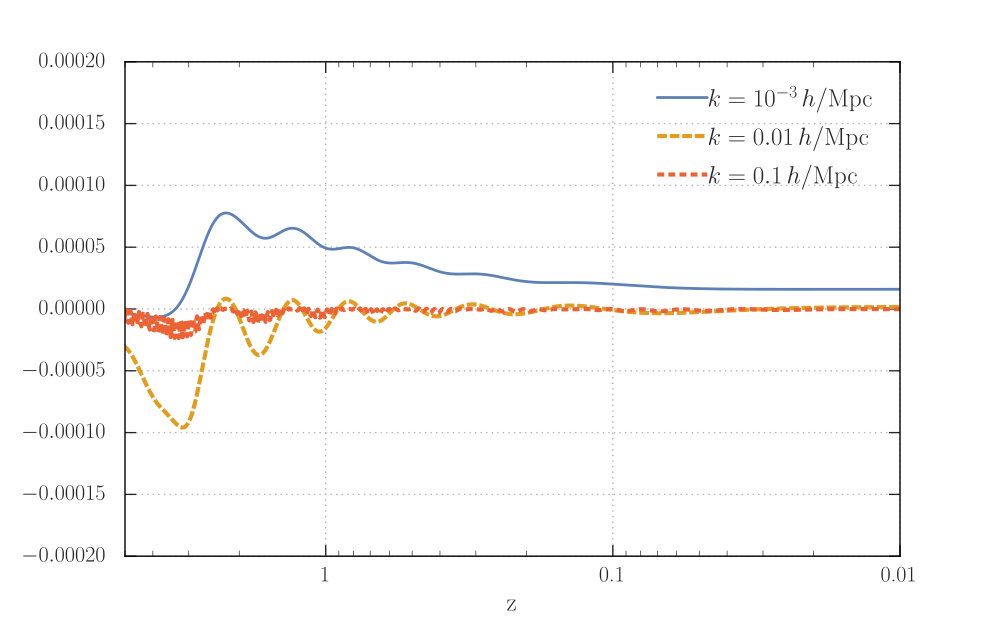}}
\subfloat{\includegraphics[width=0.49\textwidth]{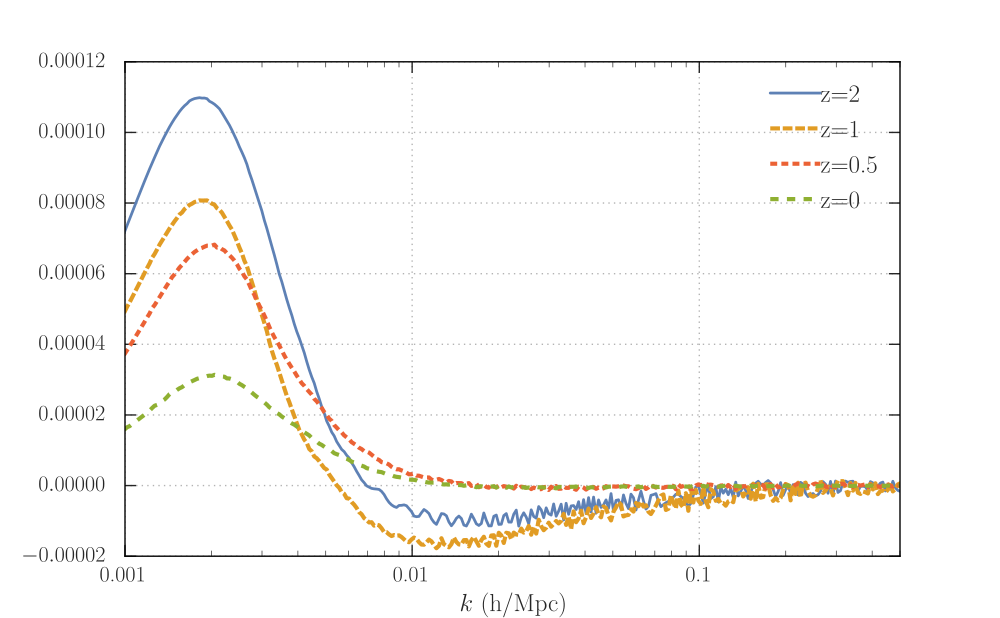}}
\caption{ {\it Top}: interacting model; {\it Bottom}: decoupled model. \emph{Left}: Fractional perturbations of the axion energy density $\delta \rho_{\rm ax} / \rho_{\rm ax}$ as a function of time, for different wavenumbers. \emph{Right}: $\delta \rho_{\rm ax} / \rho_{\rm ax}$ as a function of wavenumber, for different redshifts.} 
\label{fig:deltaax}
\end{figure*}

Here we will work in synchronous gauge, comoving with cold dark matter, defined by the line element 
\be
\rmd s^2 = - \rmd t^2 + a^2(t) \[ \delta_{ij} + \nab^{-2} \de_i \de_j (h + 6 \eta) - 2 \eta \delta_{ij} \] \rmd x^i \rmd x^j \, ,
\ee
where the gauge variable $\eta$ is given by
\begin{equation}
\frac{k^2}{a^2} \eta = \frac{H}{2} \dot{h} - \frac{3}{2} \Omega_m H^2 \delta_m - 4 \pi G ( \dot{\bar{\phi}}_J \dot{\varphi}^J + \frac{\partial V}{\partial \phi^J} \varphi^J ) \, ,
\end{equation}
and the additional condition $\delta u_m = 0$~\cite{MaBertschinger}. Here we decomposed the axion fields as $\phi^I = \bar{\phi}^I + \vphi^I$ ($I = 1,2$), splitting them into the background and the order perturbations, respectively. Then, the perturbation field equations are \cite{HwangNoh}:
\be
\ddot{\vphi}^I + 3 H \dot{\vphi}^I + \frac{k^2}{a^2} \vphi^I
+ \vphi^J \frac{\partial^2 V}{\partial \phi^I \partial \phi^J} = - \frac{\dot{h}}{2} \dot{\bar{\phi}}^I \, .
\ee
The conservation equation for matter is 
\be
\dot{\delta}_m = - \frac{\dot{h}}{2} \, .
\ee
To complete the system of independent equations, we use 
\be
\ddot{h} + 2 H \dot{h} = - 8 \pi G \( \delta \rho + 3 \delta p \)
= - 3 H^2 \Om_m \delta_m - 8 \pi G \sum_{J} \( 4 \dot{\bar{\phi}}^J \dot{\vphi}_J - 2 \frac{\de V}{\de \phi^J} \vphi^J \) \, .
\ee
We set initial conditions deep in the matter dominated era~\cite{Creminelli:2008wc}, choosing, without any loss of generality, $z=20$ as the initial time for the numerical integration of the perturbation equations.
Given an initial $\delta_m(t_{in}, \bfk)$, the last equation translates into $\dot{\delta}_m(t_{in}, \bfk) = H(t_{in}) \delta_m(t_{in}, \bfk)$.
The initial field perturbations can be set to zero, as they will quickly reach the attractor solution. Numerical results are given in Fig.
\ref{fig:deltam}, \ref{fig:deltaax}.

We find that in our model there is a small rescaling of the matter power spectrum with respect to the $\Lambda$CDM cosmology, with a suppression at higher wavenumbers, as illustrated in Fig.~\ref{fig:deltam}.
By matter perturbations here we mean those in the main component of the dark matter. Since the heavier axion is still very light, it will not cluster on the scales probed by the present galaxy surveys.

The evolution of the field perturbations is shown in Fig. \ref{fig:deltaax}, where we plot the fraction $\delta \rho_{\rm ax} / \rho_{\rm ax}$ as a function of redshift and wavenumber. An interesting observational signature of the spectra are the enhanced oscillations around a particular scale, particularly in the interacting model.
While the detailed implications for cosmological data are yet to be understood, we see that at least at present it appears that the multiaxion models do provide viable alternatives to $\Lambda$CDM.

\section{Summary}

The dark sectors of the universe could be totally separate components, referred to as dark matter, behaving as completely neutral non-relativistic billiard balls, and dark energy, modeled by a single number, the `cosmological constant'. Such a simple picture, with carefully tuned normalizations, fits the observations very well at the present time. Yet, possibilities for other options remain open. 

The `cosmic coincidences' problem, namely the uncanny similarity between the fractional contributions of very different cosmic constituents, remains an open question. Why would dark matter and dark energy influence the bending of cosmic geometry comparably at present, even though the time evolution of their effects is dramatically different today? What sets their normalizations? Could the conundrum of cosmic coincidences be resolved by some interaction between the hidden sectors of the universe, and could those also influence the visible contents? There are many very simplistic schemes attempting to address these questions by proposing to relate various sectors with interactions that could equilibrate between cosmic constituents at the level of GR+fluid sources.

At the microscopic level, the situation is not so simple. The problem rests on the fact that the evolution of different cosmic components is controlled by very different scales. If these components are allowed to interact, quantum mechanical effects communicate the presence of these scales  from one sector to another. This  basically precludes any significant interactions between heavy dark matter and light dark energy.
Specifically, the dark energy-generated long range forces between dark matter particles and the quantum radiative corrections to the dark energy mass induced by virtual dark matter particles constrain the cross-couplings to essentially zero when the masses of DM and DE are very different. 

Yet, there is a possible way out of this {\it impasse}: it involves a somewhat more complicated, but consistent, set of models of multliple axions. Some of them are light enough to be DE, some are heavy enough to be (mostly decoupled) DM, and some are too heavy to be DE today, but nevertheless
can couple strongly to DE. Setups like this emerge naturally in the construction of radiatively and non-perturbatively consistent models of field-driven cosmic acceleration, using monodromy to explain the origin of super-Planckian field displacements in effective field theory. To be clear, this does not solve the cosmic coincidence problem. However, if DE and DM are axions with comparable decay constants, and if their masses have correct values, then during inflation they will get the right initial conditions to be DM and DE today. Also, in such cases we easily find significant interactions between DE and a component of DM, with interesting observational implications.

From the low energy point of view, this is a self-consistent procedure for designing a field theory which could have a viable UV completion. But it might be more: the constructions involving many light, mixed axions could be a signal of the presence of the `axiverse', a string-theoretic realization of a low energy theory with many light axion-like fields. Such constructions make the presence of many mixed axions more natural.

We find that at the present time such models are basically degenerate with $\Lambda$CDM but have significant deviations which could be looked for in future observations. The lightest axions can mix significantly to generate such signatures.  Among the specific predictions, we note oscillations in the equation of state, sharp transitions in the Hubble parameter around a particular redshift, and possible signatures of domain walls produced by the classically-induced phase transitions of dark matter fields. These are consistent with current data but could lead to interesting small signatures at very large scales. We believe that such specific predictions warrant a dedicated analysis with a detailed comparison to data.

\section*{Acknowledgements}
We would like to thank Ed Copeland, Anne Green, Albion Lawrence, Tony Padilla, Filippo Vernizzi, George Zahariade and especially Brent Follin and Manoj Kaplinghat for useful discussions. T.H. and N.K. are supported in part by the DOE Grant DE-SC0009999.

\end{document}